\documentclass{INTERSPEECH2023}

\interspeechcameraready 

\usepackage{svg}
\usepackage{adjustbox}
\usepackage[font=small,skip=5pt]{caption}

\title{Audio-Visual Speech Separation in Noisy Environments \\ with a Lightweight Iterative Model\\}
\name{Héctor Martel$^{1,2}$
\qquad Julius Richter$^{2}$
\qquad Kai Li$^{1}$ 
\qquad Xiaolin Hu$^{1,3}$ 
\qquad Timo Gerkmann$^{2}$}

\address{$^{1}$ Department of Computer Science and Technology, 
Tsinghua University, Beijing, China \\ 
$^{2}$ Signal Processing (SP), Department of Informatics, Universität Hamburg, Hamburg, Germany \\ 
$^{3}$ Chinese Institute for Brain Research (CIBR), Beijing, China }
\email{\tt \small xlhu@tsinghua.edu.cn, timo.gerkmann@uni-hamburg.de}

\begin{document}

\maketitle
 
\begin{abstract}
We propose Audio-Visual Lightweight ITerative model (AVLIT), an effective and lightweight neural network that uses Progressive Learning (PL) to perform audio-visual speech separation in noisy environments.
To this end, we adopt the Asynchronous Fully Recurrent Convolutional Neural Network (A-FRCNN), which has shown successful results in audio-only speech separation.
Our architecture consists of an audio branch and a video branch, with iterative A-FRCNN blocks sharing weights for each modality. 
We evaluated our model in a controlled environment using the NTCD-TIMIT dataset and in-the-wild using a synthetic dataset that combines LRS3 and WHAM!. 
The experiments demonstrate the superiority of our model in both settings with respect to various audio-only and audio-visual baselines. 
Furthermore, the reduced footprint of our model makes it suitable for low resource applications. 
\end{abstract}

\section{Introduction}
\label{sec:introduction}

Understanding speech can be difficult when multiple speakers overlap, known as the \emph{cocktail party problem} \cite{cherry1953}. Speech separation has been extensively studied in the past for mixtures containing only clean speech \cite{wang2018supervised, luo2019conv}. In realistic environments, background noise added to the mixture further degrades the speech intelligibility. The suppression of such noise is accomplished by speech enhancement methods \cite{das2021fundamentals, hendriks_dft-domain_2013, gerkmann2018book_chapter}. Ultimately, the unification of both, i.e. speech separation in noisy environments, aims to bridge the gap between computer-based approaches and human performance under realistic conditions \cite{wichern2019wham}.

Traditionally, this problem has been tackled using the acoustic information alone. However, there is solid evidence that visual cues improve speech intelligibility and separation performance \cite{sumby1954visual, partan1999communication}. The principle is twofold: first, the visual information (e.g. lip movements and facial expressions) is usually not affected by the acoustic environment, and second, it has been proven that visual information is able to provide additional speech and speaker related cues \cite{mcgurk1976hearing}. The performance obtained with audio-visual methods surpasses that of audio-only methods, confirming the advantage of visual cues in adverse acoustic environments \cite{zhu2021deep, michelsanti2021overview}.

As the problem settings become closer to a real environment, there is a need to increase model capacity in favor of generalization. Large models are trending in various fields in the research community, and have achieved impressive performance at the cost of scaling up to billions of parameters. We argue that the computational demands outweigh the performance gains and make it difficult to apply these models in a practical scenario with limited hardware resources, e.g. resource-intensive audio transformer models \cite{subakan2021attention}. %
In contrast, Progressive Learning (PL) 
\cite{gao2016snr, li2019convolutional}
aims at decomposing the mapping between inputs and outputs into multiple steps. 

The key aspect of PL is the use of sub-modules with shared weights and feedback connections used in an iterative fashion. As a consequence, the sub-modules may be reduced in size with respect to a network aimed at learning the direct mapping, since the task is broken down into ``simpler'' sub-tasks and the same weights are reused multiple times.

While PL has been used for various uni-modal tasks, such as image restoration \cite{ren2019progressive}, audio-only speech enhancement \cite{gao2016snr}, or separation \cite{hu2021speech, li2022efficient}, the application of PL on a multi-modal task remains unexplored to the best of our knowledge.
To this end, we adopt the Asynchronous Fully Recurrent Convolutional Neural Network (A-FRCNN) \cite{hu2021speech} as the building block for a new multi-modal architecture.
A-FRCNN combines the principle of PL with neuroscience findings, and reports state-of-the-art (SOTA) results in audio-only speech separation.

In the experiments, we show the effectiveness of our model, called  Audio-Visual Lightweight ITerative model (AVLIT), with respect to competing audio-only and audio-visual approaches. 
Moreover, we study the impact of various design choices to find a configuration that provides both separation quality and computational efficiency.

\begin{figure*}[t]
    \centering
    \begin{tabular}{c c}
        \includegraphics[width=0.39\textwidth]{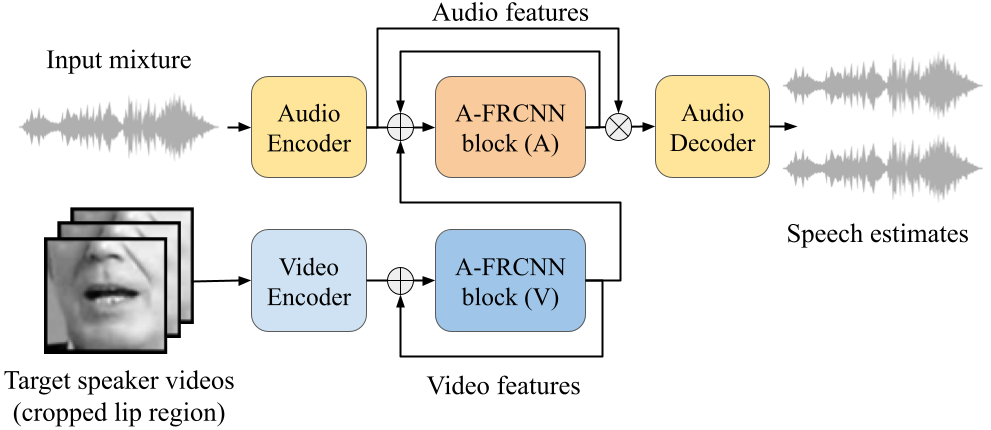}%
         &  
        \includegraphics[width=0.58\textwidth]{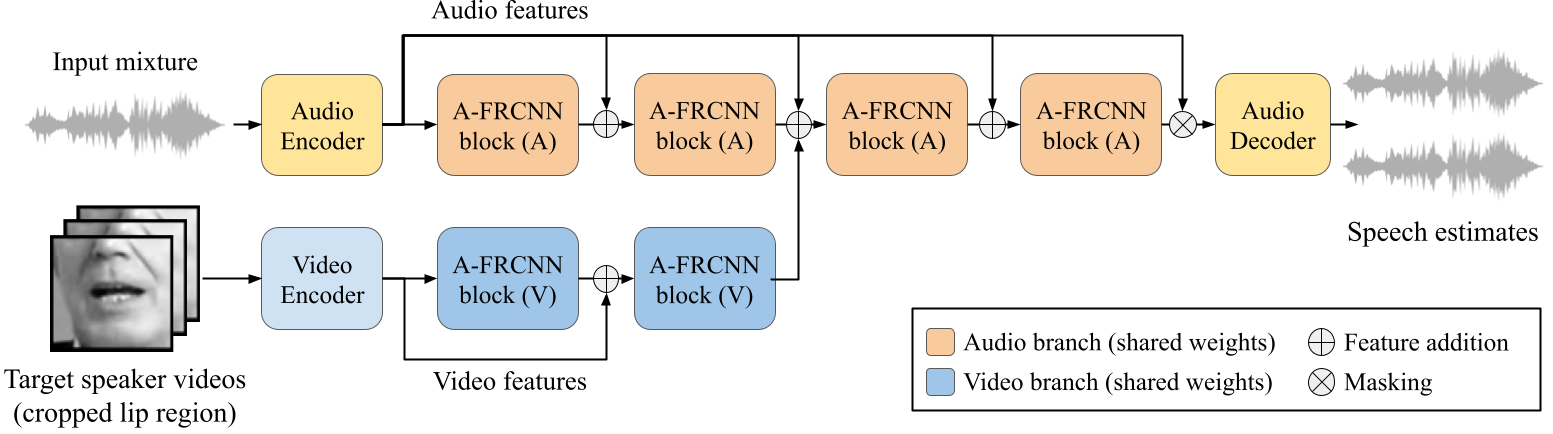}
         \\
         (A) Folded view of AVLIT. & (B) Unfolded view of AVLIT with 
         $N_A = 4, N_V = 2$, and 
         $P = \{2\}$.
    \end{tabular}
    \caption{Diagram of the proposed network architecture. The iterations are represented in folded view (A) and unfolded view (B).}
    \label{fig:av_afrcnn_architecture}
\end{figure*}

\section{Method}
\label{sec:method}

\subsection{General framework}
\label{sub:general_framework}

The task of speech separation is to extract the individual speech signals from a mixture of multiple speakers, which may also contain  background noise. The mixture, denoted by $\mathbf x \in \mathbb R^{1 \times T}$, is assumed to be additive, 
\begin{equation}
    \mathbf x = \sum_{i=1}^{M} \mathbf s_i + \mathbf n,
\end{equation}
\noindent where the time-domain signals $\mathbf s_i \in \mathbb R^{1 \times T}$ correspond to the $i$th speaker, and $\mathbf n \in \mathbb R^{1 \times T}$ to the 
background noise. 
Thus, our goal is to obtain the estimations $\hat{s}_i \in \mathbb R^{1 \times T}$ for all $i$.
$T$ denotes the number of samples, and $M$ the number of speakers. 
In the audio-visual case, the inputs
$\mathbf v_i \in \mathbb R^{F \times H \times W}$ denote a grayscale single-camera video 
corresponding to the $i$th speaker, where $F$ is the total number of frames of spatial dimensions $H \times W$.
The videos of all speakers are provided as input.

\subsection{A-FRCNN block}
\label{sub:audio_only_afrcnn}

The A-FRCNN block \cite{hu2021speech} extends the U-Net \cite{ronneberger2015unet} architecture by introducing two modifications. 
First, connections between adjacent levels are added to fuse features at different resolutions. 
Second, a \textit{delayed global fusion} 
integrates the features from all resolutions at the end.
The hyperparameters of the block are a depth of $S$ stages and $C$ channels per stage, as described in \cite{hu2021speech}. It uses $B$ input channels, which may be $B \neq C$.

\subsection{AVLIT}
\label{sub:audio_visual_afrcnn}

Our model consists of an \textit{audio branch} and a \textit{video branch}, each having A-FRCNN blocks used iteratively with shared weights. 

\textbf{Audio branch}: A representation of the time-domain audio mixture is obtained with a 1D convolutional layer in the encoder to obtain the features $\mathbf f_A \in \mathbb R^{C \times T'}$ from $\mathbf x \in \mathbb R^{1 \times T}$ ($T' \leq T$). 
Then, an A-FRCNN block is applied iteratively $N_A$ times. 
The input features $\mathbf f_A$ are added at every iteration using a skip connection from the audio encoder to the input of the next block, as in Fig. \ref{fig:av_afrcnn_architecture}.
Let $\phi(\cdot)$ denote the A-FRCNN block, $\mathbf R(i)$ the output of the block at step $i$, the iterative audio block operations are formulated as
\begin{equation}
    \mathbf R(i+1) = \phi(\mathbf R(i) + \mathbf f_A), \quad i = 0,\dots,N_A -1,
    \label{eq:audio_brach_iterative}
\end{equation}
with $\mathbf R(0)=\mathbf 0$.
The output of the last block, namely $\mathbf R(N_A) = \mathbf f'_A \in \mathbb R^{C \times T'}$, is used as mask for the input $\mathbf f_A$. The result $\mathbf f_A \odot \mathbf f'_A$ is decoded into $M$ channels using a 1D transposed convolution. 
Thus, we obtain the speech estimations $\{ \hat{\mathbf s}_1, ..., \hat{\mathbf s}_M \} \in \mathbb R^{1 \times T}$.

\textbf{Video branch}: Frame-level embeddings for all speakers $\mathbf f_V \in \mathbb R^{M \times C' \times F} $ are obtained from 
$\mathbf v_i $
using the latent representation with dimension $C'$ of a pre-trained single-frame autoencoder.
Then, an A-FRCNN block is applied iteratively $N_V$ times.
Finally, $\mathbf f_V$ is fixed to match $\mathbf f_A$ by applying $1 \times 1$ convolutions to the channels and nearest neighbor interpolation to the temporal dimension, obtaining $\mathbf f'_V \in \mathbb R^{C \times T'}$ (without information loss since $F \ll T'$).

\textbf{Modality fusion}: The features $\mathbf f'_V$ processed by the video branch are injected directly at specific positions within the iteration of the audio branch, as it can be seen in Fig. \ref{fig:av_afrcnn_architecture} (B). 
These positions are defined by a set $P$, with elements in the interval $[0,\dots,N_A -1]$. 
With this modification, the iterative process from Eq. \eqref{eq:audio_brach_iterative} becomes

 \begin{equation}
  \mathbf R(i+1) = \begin{cases}
        \phi(\mathbf R(i) + \mathbf f_A + \mathbf f'_V), &\quad i \in P
        \\
        \phi(\mathbf R(i) + \mathbf f_A), &\quad i \notin P,
        \end{cases}
        \label{eq:av_fusion}
 \end{equation}
for $i = 0, \dots, N_A -1$ and $\mathbf R(0)= \mathbf 0$.
 The operation to combine $\mathbf f_A$ and $\mathbf f'_V$ is addition. 
 We found no significant performance differences between addition, product or concatenation. 

 \textbf{Complexity analysis}: Let $N = N_A + N_V$. 
 The time complexity is $O(N)$, since the operations increase linearly with the total number of blocks, as per Fig.\ref{fig:av_afrcnn_architecture} (B). However, the space complexity is $O(1)$, because the parameters are shared within each branch 
 for any choice of 
 $N_A$ and $N_V$, as per Fig.\ref{fig:av_afrcnn_architecture} (A).

 \section{Experimental settings}
\label{sec:experimental_settings}

\subsection{Datasets}
\label{sub:dataset}

We use NTCD-TIMIT for a controlled environment and LRS3+ WHAM! for data in the wild.
Both datasets are pre-processed to use the lip region of the video recordings as done in \cite{afouras2018deep}. 

\textbf{NTCD-TIMIT~\cite{abdelaziz2017ntcd}} 
was proposed for noise-robust audio-visual speech recognition for a single speaker. We generate mixtures of 
2 speakers plus noise by
selecting an additional speaker randomly from the same set without overlap in speaker identity or utterance content. 
The SNR is 0 dB for clean speech and uniformly sampled in the range $[-5, 20]$ dB for the noise. 
The data splits have 39, 8 and 9 speakers, with 5 hours for training, 1 hour for validation, and 1 hour for testing, respectively. 

\textbf{LRS3~\cite{afouras2018lrs3} + WHAM!~\cite{wichern2019wham}} combines the audio-visual data from LRS3 and the noise data from WHAM!. We generated synthetic mixtures containing 
2 
speakers plus noise. The SNR is uniformly sampled in $[-5, 5]$ dB for clean speech and $[-6, 3]$ dB for noise, following the mixing procedure originally described in WHAM!. We obtained 28 hours for training, 3 hours for validation, and 2 hours for testing.

\begin{table*}[t]
    \caption{Comparison with baseline models on NTCD-TIMIT and LRS3+WHAM! datasets. We report instrumental audio quality and intelligibility metrics for each dataset using mixtures of 2 speakers plus noise.
    }
    \label{tab:baseline_quality_comparison}
    \centering
    \scalebox{0.78}{
        \begin{tabular}{l | c c | r r r | r r r }
	\hline
	\multicolumn{1}{c}{} & \multicolumn{2}{c}{Inputs} & \multicolumn{3}{c}{NTCD-TIMIT} & \multicolumn{3}{c}{LRS3+WHAM!} 
	\\
	Model                                 & A            & V            & PESQ $\uparrow$ & ESTOI $\uparrow$ & SI-SDRi $\uparrow$ & PESQ $\uparrow$ & ESTOI $\uparrow$              & SI-SDRi $\uparrow$ \\
	\hline
	Unprocessed                           & --           & --           & 1.19            & 0.33             & --                 & 1.05            & 0.34                          & --                 \\
	\hline
	ConvTasNet \cite{luo2019conv}         & $\checkmark$ & $\times$     & 1.35            & 0.38             & 8.76               & 1.24            & 0.51                          & 9.66               \\
	DPRNN \cite{luo2020dual}              & $\checkmark$ & $\times$     & 1.32            & 0.39             & 9.31               & 1.40            & 0.45                          & 10.93              \\
	A-FRCNN-2 \cite{hu2021speech}         & $\checkmark$ & $\times$     & 1.22            & 0.21             & 5.94               & 1.18            & 0.46                          & 6.48               \\
	A-FRCNN-4 \cite{hu2021speech}         & $\checkmark$ & $\times$     & 1.24            & 0.26             & 6.61               & 1.23            & 0.46                          & 7.32               \\
	A-FRCNN-8 \cite{hu2021speech}         & $\checkmark$ & $\times$     & 1.30            & 0.31             & 6.92               & 1.25            & 0.49                          & 9.21               \\
	\hline
	AVConvTasNet \cite{wu2019time}        & $\checkmark$ & $\checkmark$ & 1.33            & 0.40             & 9.02               & 1.29            & 0.60                          & 6.21               \\
	LAVSE \cite{Chuang2020}               & $\checkmark$ & $\checkmark$ & 1.31            & 0.37             & 6.22               & 1.24            & 0.50                          & 5.59               \\
	L2L \cite{ephrat2018looking}          & $\checkmark$ & $\checkmark$ & 1.23            & 0.26             & 3.36               & 1.16            & 0.51                          & 7.60               \\
	VisualVoice \cite{gao2021visualvoice} & $\checkmark$ & $\checkmark$ & \textbf{1.45}   & 0.43             & 10.04              & 1.48            & 0.63                          & 11.87              \\
	\hline
	AVLIT-2 (Ours)                        & $\checkmark$ & $\checkmark$ & 1.36            & 0.40             & 9.05               & 1.30            & 0.55                          & 9.10               \\
	AVLIT-4 (Ours)                        & $\checkmark$ & $\checkmark$ & 1.38            & 0.42             & 10.15              & 1.41            & 0.62                          & 10.99              \\
	AVLIT-8 (Ours)                        & $\checkmark$ & $\checkmark$ & 1.43            & \textbf{0.45}    & \textbf{11.00}     & \textbf{1.52}   & \textbf{0.68}                 & \textbf{12.42}     \\
	\hline
\end{tabular}
    }
\end{table*}

\subsection{Evaluation}
\label{sub:evaluation}

We evaluate our method with respect to the separation quality and the computational efficiency.

\textbf{Separation quality}: Instrumental evaluation metrics such as Perceptual Evaluation of Speech Quality (PESQ) \cite{rix2001pesq},
Extended Short-Time Objective Intelligibility (ESTOI) \cite{jensen2016algorithm}, and Scale-Invariant Signal-to-Distortion Ratio (SI-SDRi) \cite{le2019sdr} are used to respectively measure the speech quality, intelligibility, and overall distortions of the speech estimations. 

\textbf{Computational efficiency}: The efficiency is compared using Multiply-Accumulate operations (MACs), number of parameters, inference times on CPU and GPU, and memory footprint during training and inference 
(computed with the \textit{ptflops} package\footnote{\url{https://github.com/sovrasov/flops-counter.pytorch}}). 
Inference time is measured with 2s of audio at 16 kHz sampling rate and averaged over 100 trials on the CPU and GPU models specified in Sec. \ref{sub:implementation_details}. 

\subsection{Implementation details}
\label{sub:implementation_details}

The hyperparameters used in the architecture and experimental settings are detailed below:

\textbf{Audio encoder/decoder}: The audio encoder and decoder employ a 1D convolution and a 1D transposed convolution, respectively. The kernel size is set to $K = 40$ (2.5 ms) with a stride of 20 (1.25 ms), and $C = 512$ output channels. 

\textbf{Video encoder}: The video encoder takes the video frames of the cropped lip region, resized to 64$\times$64 px, and generates frame-level embeddings using a 4-layer convolutional autoencoder (AE) with bottleneck of size $C'=1024$. 
Each layer performs a 2D convolution with kernel size 2 and stride 2, followed by a LeakyReLU activation with slope 0.3. 
The AE is trained for each dataset by minimizing the mean squared error between the input frame and its reconstruction. The encoder is used in our model with frozen weights.

\textbf{Audio branch}: 
The audio block uses $C_A = 512$, $B_A = 128$ and $S_A = 5$.
As a result, it has 4.9 M parameters. 
The number of blocks of the audio branch is set to $N_A = [2,4,8]$ and we denote our model as AVLIT-$N_A$.

\textbf{Video branch}: 
The video block has a reduced size since $F \ll T'$, i.e. the video features have a significantly lower rate than the audio features. 
The block uses $C_V = 128$, $B_V = 128$ and $S_V = 5$, resulting in 0.35 M parameters.
We found no practical differences in speech separation quality when the size is reduced, while it makes the model significantly lighter. 
The number of blocks of the video branch is set to $N_V = N_A / 2$, which we found to work well experimentally.

\textbf{Training procedure}: All models were trained for 100 epochs on 4-second utterances for NTCD-TIMIT and 2-second utterances for LRS3 + WHAM! at 16 kHz sampling rate. We used a batch size of 16, AdamW optimizer \cite{loshchilov2018decoupled} with a learning rate of $1 \times 10^{-3}$, weight decay of $1 \times 10^{-1}$, and a step learning rate schedule with a factor of $1/3$ applied every 25 epochs.
The training objective is the negative SI-SDR \cite{le2019sdr}. 
The speaker assignment in audio-visual models preserves the order of the visual inputs, while audio-only models use permutation invariant training \cite{yu2017permutation}.

\textbf{Hardware configuration}: All experiments were conducted on a server with Intel(R) Xeon(R) Silver 4210 CPU @ 2.20 GHz and 8 $\times$ GeForce RTX 2080 Ti 11GB. 

\textbf{Official code}: The Pytorch implementation of the model is available\footnote{\url{https://github.com/hmartelb/avlit}}.
The A-FRCNN block is adopted from its original implementation\footnote{\url{https://cslikai.cn/project/AFRCNN}}. This project is MIT Licensed.

\begin{table*}[t]
    \caption{Comparison with baseline models in terms of computational efficiency. We report model complexity, inference times on both CPU and GPU, and total memory allocated during training and inference. An asterisk (*) denotes overhead due to one or more visual frontends, although not accounted here. All numbers are obtained with audios of length 2s at 16kHz sampling rate and batch size of 1.}
    \label{tab:baseline_efficiency_comparison}
    \centering
    \scalebox{0.78}{
    \begin{tabular}{l | c c | r r | r r | r r }
	\hline
	\multicolumn{1}{c}{} & \multicolumn{2}{c}{Inputs} & \multicolumn{2}{c}{Model complexity} & \multicolumn{2}{c}{Inference time} & \multicolumn{2}{c}{Memory footprint} \\
	Model                                 & A            & V            & MACs (G) $\downarrow$ & Params (M) $\downarrow$ & CPU (s) $\downarrow$ & GPU (s) $\downarrow$ & Train (MB) $\downarrow$ & Inference (MB) $\downarrow$ \\
	\hline
	ConvTasNet \cite{luo2019conv}         & $\checkmark$ & $\times$     & 28.03                 & 3.50                    & 0.14                         & \textbf{0.02}                & 2172.66                 & 20.23                       \\
	DPRNN \cite{luo2020dual}              & $\checkmark$ & $\times$     & 168.77                & \textbf{2.62}           & 4.68                         & 0.46                         & 7484.10                 & \textbf{14.61}              \\
	A-FRCNN-2 \cite{hu2021speech}         & $\checkmark$ & $\times$     & \textbf{10.31}        & 5.14                    & \textbf{0.10}                & 0.03                         & \textbf{847.24}         & 21.55                       \\
	A-FRCNN-4 \cite{hu2021speech}         & $\checkmark$ & $\times$     & 18.96                 & 5.14                    & 0.18                         & 0.04                         & 1535.25                 & 21.55                       \\
	A-FRCNN-8 \cite{hu2021speech}         & $\checkmark$ & $\times$     & 36.27                 & 5.14                    & 0.38                         & 0.06                         & 2909.77                 & 21.55                       \\
	\hline
	AVConvTasNet \cite{wu2019time}        & $\checkmark$ & $\checkmark$ & 43.57 *                & 16.45                   & 0.43                         & 0.07                         & 2474.98 *               & 117.05                      \\
	LAVSE \cite{Chuang2020}               & $\checkmark$ & $\checkmark$ & 206.51                & 32.89                   & 0.55                         & 0.05                         & 2386.82                 & 178.61                      \\
	L2L \cite{ephrat2018looking}          & $\checkmark$ & $\checkmark$ & 93.58 *                 & 17.40                   & 0.57                         & 0.04                         & 834.68 *                & 116.83                      \\
	VisualVoice \cite{gao2021visualvoice} & $\checkmark$ & $\checkmark$ & 19.70 *                 & 77.75                   & 1.98                         & 0.20                         & 5346.52 *               & 313.74                      \\
	\hline
	AVLIT-2 (Ours)                        & $\checkmark$ & $\checkmark$ & 10.37                 & 5.75                    & 0.11                         & 0.04                         & 881.92                  & 23.97                       \\
	AVLIT-4 (Ours)                        & $\checkmark$ & $\checkmark$ & 19.03                 & 5.75                    & 0.20                         & 0.05                         & 1570.98                 & 23.97                       \\
	AVLIT-8 (Ours)                        & $\checkmark$ & $\checkmark$ & 36.35                 & 5.75                    & 0.48                         & 0.08                         & 2949.11                 & 23.97                       \\
	\hline
\end{tabular}
    }
\end{table*}

\section{Results}
\label{sec:results}

\subsection{Comparison with existing models}

We select ConvTasNet \cite{luo2019conv}, DualPathRNN (DPRNN) \cite{luo2020dual}, and A-FRCNN \cite{hu2021speech} as audio-only baselines and AVConvTasNet \cite{wu2019time}, LAVSE \cite{Chuang2020}, Looking to Listen (L2L) \cite{ephrat2018looking}, and VisualVoice \cite{gao2021visualvoice} as audio-visual baselines. 
All models are trained with our settings using the official code when available, or the implementation from the Asteroid \footnote{\url{https://github.com/asteroid-team/asteroid}} otherwise. 
The separation quality results are shown in Table \ref{tab:baseline_quality_comparison} and computational efficiency results in Table \ref{tab:baseline_efficiency_comparison}.

\textbf{Separation quality}: In Table 1, we observe that AVLIT-8 is the best performing model overall for both datasets, providing consistent improvements over the baseline methods. 
The margin is larger in LRS3+WHAM!, which contains more challenging conditions but also has more training data than NTCD-TIMIT. 
AVLIT-8 outperforms VisualVoice, the best audio-visual baseline, by rougly 1 dB and 0.5 dB SI-SDRi in the two datasets, respectively. 
AVLIT-8 and AVLIT-4 also yield better performance than DPRNN, the best audio-only baseline. 
Finally, AVLIT achieves improvements of over 3 dB SI-SDRi w.r.t. the A-FRCNN models of equivalent size, showing that the video modality is integrated effectively in our architecture.  

\textbf{Computational efficiency}: In Table 2, it can be observed that the model complexity of AVLIT is reduced significantly w.r.t. other audio-visual models. 
MACs increase linearly and the number of parameters are constant in AVLIT as a consequence of its iterative design.
In both metrics AVLIT outperforms AVConvTasNet, the audio-visual baseline with the least complexity, and it is comparable to audio-only models. 
In addition, AVLIT runs faster inference and has a smaller memory footprint than VisualVoice and DPRNN, which are the best performing baselines in terms of audio quality. 
The memory footprint is comparable to running audio-only models, requiring just 23.97 MB, which is 4.87 times less than any other audio-visual baseline (L2L requires 116.83 MB). 
For reference, VisualVoice requires 313.74 MB (13.08 times AVLIT) and needs 1.98 s to run inference on CPU (4.12 times slower than AVLIT-8).
While DPRNN requires only 14.61 MB, the recurrent nature of the model makes its inference time significantly slower, with 4.68 s on CPU (9.75 times slower than AVLIT-8).
In summary, the computational efficiency metrics show that our model is lightweight and that it excels at inference. 

\subsection{Ablation study}

We select LRS3 + WHAM! to conduct an ablation study of our model. 
We investigate the fusion position, the capacity of each branch, the effect of the number of iterations of the video branch ($N_V$), and the role of different video features. 
The number of audio blocks is set to $N_A = 4$ in this section.

\textbf{Fusion position}: In Eq. \eqref{eq:av_fusion}, we introduced the set of positions $P$ at which the video features are combined with the audio features. The results for different choices of $P$ are shown in Table \ref{tab:fusion_position_study}.
In this comparison, \textit{Early} fusion has a positive effect on the outcome, consistent with a previous study \cite{kuo2022inferring}.
We hypothesize that the audio-visual fusion of low-level features is more effective and that \textit{Early} and \textit{All} better respect the regular structure of the iterative network. %

\begin{table}[ht]
    \caption{Study of the fusion position.}
    \label{tab:fusion_position_study}
    \centering
        \scalebox{0.80}{
            \begin{tabular}{lcrrr}
            \hline
            Fusion pos.                   & $P$ & PESQ $\uparrow$            & ESTOI $\uparrow$          & SI-SDRi $\uparrow$         \\
            \hline
            Early & $\{0\}$ &    \textbf{1.41} &  \textbf{0.62} &   \textbf{10.99} \\
            Middle & $\{2\}$ &    1.30 &  0.56 &    9.15 \\
            Late & $\{3\}$ &    1.29 &  0.55 &    8.88 \\
            All & $\{0, 1, 2, 3\}$ &    1.32 &  0.58 &    9.76 \\
            \hline
            \end{tabular}
        }
\end{table}

\textbf{Branch capacity}: 
We investigate the effect of decreasing $C_A$ and $C_V$, aiming to find more efficient choices without sacrificing separation quality. We use the original values of $B=128$ as lower bound and $C=512$ as upper bound. 
The results are shown in Table \ref{tab:branch_capacity_study}.
There is degradation by reducing the value of $C_A$, but the differences are minimal when reducing $C_V$. 
This suggests that $C_A = 512$ and $C_V = 128$ provide a good compromise between separation quality and efficiency. 
We hypothesize that audio needs more capacity than video because this task requires audio outputs, while the video is used as side information.

\begin{table}[ht]
    \centering
    \caption{Study of the branch capacity, changing the number of channels for each modality.}
    \scalebox{0.78}{
    \begin{tabular}{c c r r r r}
         \hline
         $C_A$ & $C_V$ & Params (M) $\downarrow$ & PESQ $\uparrow$ & ESTOI $\uparrow$ & SI-SDRi $\uparrow$\\
         \hline
         128 & 128 & 1.01 & 1.28 & 0.55 & 8.99 \\
         256 & 128 & 2.00 & 1.35 & 0.59 & 10.19 \\
         512 & 128 & 5.75 & \textbf{1.41} & \textbf{0.62} & 10.99 \\
         512 & 256 & 6.68 & \textbf{1.41} & \textbf{0.62} & 11.00 \\
         512 & 512 & 10.30 & \textbf{1.41} & \textbf{0.62} & \textbf{11.02} \\
         \hline
    \end{tabular}
    }
    \label{tab:branch_capacity_study}
\end{table}

\textbf{Length of the video branch}: The number of audio iterations, $N_A$ is fixed to 4 and $N_V$ takes the values $[0,1,2,4]$. The results are shown in Table \ref{tab:visual_length_study}. Although performance differences are slight, we observe that $N_V = 2$ obtained the best results. 
Thus, $N_V = N_A / 2$, as per Sec. \ref{sub:implementation_details}, is reasonable.
\begin{table}[ht]
    \caption{Study of the length of the visual branch.}
    \label{tab:visual_length_study}
    \centering
    \scalebox{0.80}{
        \begin{tabular}{c c r r r }
            \hline
             $N_V$              & PESQ $\uparrow$                     & ESTOI $\uparrow$                   & SI-SDRi $\uparrow$                   \\
            \hline
             0                              & 1.38           & 0.61         & 10.67       \\
             1                              & 1.39          & \textbf{0.62} & 10.85    \\
             2                              & \textbf{1.41} & \textbf{0.62} & \textbf{10.99} \\
             4                              & 1.40           & \textbf{0.62} & 10.88           \\
            \hline
        \end{tabular}
    }
\end{table}

\textbf{Video features}: 
We compare embeddings from different video encoders.
FaRL \cite{zheng2022general} is recent SOTA in face representation learning. 
We compute face embeddings of each speaker from the original raw videos. 
AE reconstructs single frames from the cropped lip regions. 
Similarly, LipReadingTCN \cite{ma2021towards}, a successful lip reading model, is used to compute embeddings of linguistic context. 
We train AVLIT-4 replacing only the video encoder. 
The results shown in Table \ref{tab:autoencoder_vs_lipreading_results}.
In terms of separation quality, LipReadingTCN yields the best results followed by AE with minor differences. 
Considering efficiency, however, LipReadingTCN is significantly more expensive than AE since they have 32M and 5.5k parameters, respectively. 
To make our model lightweight without sacrificing separation quality, we find AE to be a sensible choice. 

\begin{table}[ht]
    \caption{Comparison between different video encoders.}
    \label{tab:autoencoder_vs_lipreading_results}
    \centering
    \scalebox{0.78}{
        \begin{tabular}{l r r r r }
            \hline
             Video encoder & Params (M) $\downarrow$                   & PESQ $\uparrow$            & ESTOI $\uparrow$          & SI-SDRi $\uparrow$         \\
            \hline
             FaRL \cite{zheng2022general}                   & 87 & 1.40 & 0.61 & 10.76 \\
             AE \cite{Chuang2020} &     0.005              & 1.41          & \textbf{0.62}          & 10.99          \\
             LipReadingTCN \cite{ma2021towards} &     32             & \textbf{1.42}          & \textbf{0.62}          & \textbf{11.05}          \\
            \hline
        \end{tabular}
    }
\end{table}

\section{Conclusion}
\label{sec:conclusion}

In this paper, we proposed
an iterative 
model for audio-visual speech separation that uses PL in audio and video branches.
Its design is based on the A-FRCNN block, which we have used for both modalities to keep a unified architecture.

The resulting model, called AVLIT, outperformed existing baselines in terms of audio quality and computational efficiency. 
We investigated the impact of various design choices to achieve a good compromise in both aspects.

As for the limitations of our work, the behavior of AVLIT in reverberant environments is unknown. 
Also, the speakers must be visible in the video frame since we assume no occlusions.

\section{Acknowledgements}

 This work was supported in part by the National Key Researchand Development Program of China (No. 2017YFA0700904), the German Research Foundation (DFG) in the transregio project Crossmodal Learning (TRR 169), the National Natural Science Foundation of China (Nos. 62061136001 and U19B2034) and the Tsinghua-Toyota Joint Research Fund. Martel is also supported by a grant from “la Caixa” Banking Foundation (ID100010434) code LCF/BQ/AA19/11720035.

\bibliographystyle{IEEEtran}
\bibliography{main}

\begin{thebibliography}{10}
\providecommand{\url}[1]{#1}
\csname url@samestyle\endcsname
\providecommand{\newblock}{\relax}
\providecommand{\bibinfo}[2]{#2}
\providecommand{\BIBentrySTDinterwordspacing}{\spaceskip=0pt\relax}
\providecommand{\BIBentryALTinterwordstretchfactor}{4}
\providecommand{\BIBentryALTinterwordspacing}{\spaceskip=\fontdimen2\font plus
\BIBentryALTinterwordstretchfactor\fontdimen3\font minus
  \fontdimen4\font\relax}
\providecommand{\BIBforeignlanguage}[2]{{%
\expandafter\ifx\csname l@#1\endcsname\relax
\typeout{** WARNING: IEEEtran.bst: No hyphenation pattern has been}%
\typeout{** loaded for the language `#1'. Using the pattern for}%
\typeout{** the default language instead.}%
\else
\language=\csname l@#1\endcsname
\fi
#2}}
\providecommand{\BIBdecl}{\relax}
\BIBdecl

\bibitem{cherry1953}
E.~C. Cherry, ``Some experiments on the recognition of speech, with one and
  with two ears,'' \emph{The Journal of the Acoustical Society of America},
  vol.~25, no.~5, pp. 975--979, 1953.

\bibitem{wang2018supervised}
D.~Wang and J.~Chen, ``Supervised speech separation based on deep learning: An
  overview,'' \emph{IEEE/ACM Transactions on Audio, Speech, and Language
  Processing}, vol.~26, no.~10, pp. 1702--1726, 2018.

\bibitem{luo2019conv}
Y.~Luo and N.~Mesgarani, ``Conv-{TasNet}: Surpassing ideal time--frequency
  magnitude masking for speech separation,'' \emph{IEEE/ACM Transactions on
  Audio, Speech, and Language Processing}, vol.~27, no.~8, pp. 1256--1266,
  2019.

\bibitem{das2021fundamentals}
N.~Das, S.~Chakraborty, J.~Chaki, N.~Padhy, and N.~Dey, ``Fundamentals, present
  and future perspectives of speech enhancement,'' \emph{Int. Journal of Speech
  Technology}, vol.~24, no.~4, pp. 883--901, 2021.

\bibitem{hendriks_dft-domain_2013}
R.~C. Hendriks, T.~Gerkmann, and J.~Jensen,
  \emph{\BIBforeignlanguage{eng}{{{DFT}}-{{Domain Based Single}}-{{Microphone
  Noise Reduction}} for {{Speech Enhancement}}: A {{Survey}} of the
  {{State}}-of-the-{{Art}}}}, ser. Synthesis Lectures on Speech and Audio
  Processing.\hskip 1em plus 0.5em minus 0.4em\relax {Williston, VT}: {Morgan
  \& Claypool}, 2013, no.~11.

\bibitem{gerkmann2018book_chapter}
T.~Gerkmann and E.~Vincent, ``Spectral masking and filtering,'' in \emph{Audio
  Source Separation and Speech Enhancement}, E.~Vincent, T.~Virtanen, and
  S.~Gannot, Eds.\hskip 1em plus 0.5em minus 0.4em\relax John Wiley \& Sons,
  2018.

\bibitem{wichern2019wham}
G.~Wichern, J.~Antognini, M.~Flynn, L.~R. Zhu, E.~McQuinn, D.~Crow, E.~Manilow,
  and J.~Le~Roux, ``{WHAM!}: Extending speech separation to noisy
  environments,'' \emph{Proc. Interspeech 2019}, pp. 1368--1372, 2019.

\bibitem{sumby1954visual}
W.~H. Sumby and I.~Pollack, ``Visual contribution to speech intelligibility in
  noise,'' \emph{The Journal of the Acoustical Society of America}, vol.~26,
  no.~2, pp. 212--215, 1954.

\bibitem{partan1999communication}
S.~Partan and P.~Marler, ``Communication goes multimodal,'' \emph{Science},
  vol. 283, no. 5406, pp. 1272--1273, 1999.

\bibitem{mcgurk1976hearing}
H.~McGurk and J.~MacDonald, ``Hearing lips and seeing voices,'' \emph{Nature},
  vol. 264, no. 5588, pp. 746--748, 1976.

\bibitem{zhu2021deep}
H.~Zhu, M.-D. Luo, R.~Wang, A.-H. Zheng, and R.~He, ``Deep audio-visual
  learning: A survey,'' \emph{Int. Journal of Automation and Computing},
  vol.~18, no.~3, pp. 351--376, 2021.

\bibitem{michelsanti2021overview}
D.~Michelsanti, Z.-H. Tan, S.-X. Zhang, Y.~Xu, M.~Yu, D.~Yu, and J.~Jensen,
  ``An overview of deep-learning-based audio-visual speech enhancement and
  separation,'' \emph{IEEE/ACM Transactions on Audio, Speech, and Language
  Processing}, vol.~29, 2021.

\bibitem{subakan2021attention}
C.~Subakan, M.~Ravanelli, S.~Cornell, M.~Bronzi, and J.~Zhong, ``Attention is
  all you need in speech separation,'' in \emph{IEEE International Conference
  on Acoustics, Speech and Signal Processing (ICASSP)}, 2021, pp. 21--25.

\bibitem{gao2016snr}
T.~Gao, J.~Du, L.-R. Dai, and C.-H. Lee, ``{SNR}-based progressive learning of
  deep neural network for speech enhancement.'' in \emph{Interspeech}, 2016,
  pp. 3713--3717.

\bibitem{li2019convolutional}
A.~Li, M.~Yuan, C.~Zheng, and X.~Li, ``Speech enhancement using progressive
  learning-based convolutional recurrent neural network,'' \emph{Applied
  Acoustics}, vol. 166, p. 107347, 2020.

\bibitem{ren2019progressive}
D.~Ren, W.~Zuo, Q.~Hu, P.~Zhu, and D.~Meng, ``Progressive image deraining
  networks: A better and simpler baseline,'' in \emph{Proceedings of the
  IEEE/CVF Conference on Computer Vision and Pattern Recognition}, 2019, pp.
  3937--3946.

\bibitem{hu2021speech}
X.~Hu, K.~Li, W.~Zhang, Y.~Luo, J.-M. Lemercier, and T.~Gerkmann, ``Speech
  separation using an asynchronous fully recurrent convolutional neural
  network,'' in \emph{Advances in Neural Information Processing Systems},
  vol.~34, 2021, pp. 22\,509--22\,522.

\bibitem{li2022efficient}
K.~Li, R.~Yang, and X.~Hu, ``An efficient encoder-decoder architecture with
  top-down attention for speech separation,'' in \emph{ICLR}, 2023.

\bibitem{ronneberger2015unet}
O.~Ronneberger, P.~Fischer, and T.~Brox, ``{U-Net}: Convolutional networks for
  biomedical image segmentation,'' in \emph{International Conference on Medical
  image computing and computer-assisted intervention}.\hskip 1em plus 0.5em
  minus 0.4em\relax Springer, 2015, pp. 234--241.

\bibitem{afouras2018deep}
T.~Afouras, J.~S. Chung, A.~Senior, O.~Vinyals, and A.~Zisserman, ``Deep
  audio-visual speech recognition,'' \emph{IEEE Transactions on Pattern
  Analysis and Machine Intelligence}, 2018.

\bibitem{abdelaziz2017ntcd}
A.~H. Abdelaziz, ``{NTCD-TIMIT}: A new database and baseline for noise-robust
  audio-visual speech recognition,'' \emph{Proc. Interspeech 2017}, pp.
  3752--3756, 2017.

\bibitem{afouras2018lrs3}
T.~Afouras, J.~S. Chung, and A.~Zisserman, ``{LRS3-TED}: a large-scale dataset
  for visual speech recognition,'' \emph{arXiv preprint arXiv:1809.00496},
  2018.

\bibitem{luo2020dual}
Y.~Luo, Z.~Chen, and T.~Yoshioka, ``Dual-path {RNN}: efficient long sequence
  modeling for time-domain single-channel speech separation,'' in \emph{IEEE
  International Conference on Acoustics, Speech and Signal Processing
  (ICASSP)}, 2020, pp. 46--50.

\bibitem{wu2019time}
J.~Wu, Y.~Xu, S.-X. Zhang, L.-W. Chen, M.~Yu, L.~Xie, and D.~Yu, ``Time domain
  audio visual speech separation,'' in \emph{IEEE Automatic Speech Recognition
  and Understanding Workshop (ASRU)}, 2019, pp. 667--673.

\bibitem{Chuang2020}
S.-Y. Chuang, Y.~Tsao, C.-C. Lo, and H.-M. Wang, ``Lite audio-visual speech
  enhancement,'' in \emph{Proc. Interspeech 2020}, 2020, pp. 1131--1135.

\bibitem{ephrat2018looking}
A.~Ephrat, I.~Mosseri, O.~Lang, T.~Dekel, K.~Wilson, A.~Hassidim, W.~T.
  Freeman, and M.~Rubinstein, ``Looking to listen at the cocktail party: a
  speaker-independent audio-visual model for speech separation,'' \emph{ACM
  Transactions on Graphics (TOG)}, vol.~37, no.~4, pp. 1--11, 2018.

\bibitem{gao2021visualvoice}
R.~Gao and K.~Grauman, ``Visualvoice: Audio-visual speech separation with
  cross-modal consistency,'' in \emph{2021 IEEE/CVF Conference on Computer
  Vision and Pattern Recognition (CVPR)}.\hskip 1em plus 0.5em minus
  0.4em\relax IEEE, 2021, pp. 15\,490--15\,500.

\bibitem{rix2001pesq}
A.~Rix, J.~Beerends, M.~Hollier, and A.~Hekstra, ``Perceptual evaluation of
  speech quality ({PESQ})-a new method for speech quality assessment of
  telephone networks and codecs,'' in \emph{IEEE Int. Conference on Acoustics,
  Speech, and Signal Processing. Proceedings}, vol.~2, 2001, pp. 749--752
  vol.2.

\bibitem{jensen2016algorithm}
J.~Jensen and C.~H. Taal, ``An algorithm for predicting the intelligibility of
  speech masked by modulated noise maskers,'' \emph{IEEE/ACM Transactions on
  Audio, Speech, and Language Processing}, vol.~24, no.~11, pp. 2009--2022,
  2016.

\bibitem{le2019sdr}
J.~Le~Roux, S.~Wisdom, H.~Erdogan, and J.~R. Hershey, ``{SDR}--half-baked or
  well done?'' in \emph{IEEE International Conference on Acoustics, Speech and
  Signal Processing (ICASSP)}, 2019, pp. 626--630.

\bibitem{loshchilov2018decoupled}
I.~Loshchilov and F.~Hutter, ``Decoupled weight decay regularization,'' in
  \emph{International Conference on Learning Representations}, 2018.

\bibitem{yu2017permutation}
D.~Yu, M.~Kolb{\ae}k, Z.-H. Tan, and J.~Jensen, ``Permutation invariant
  training of deep models for speaker-independent multi-talker speech
  separation,'' in \emph{IEEE International Conference on Acoustics, Speech and
  Signal Processing (ICASSP)}, 2017, pp. 241--245.

\bibitem{kuo2022inferring}
T.-Y. Kuo, Y.~Liao, K.~Li, B.~Hong, and X.~Hu, ``Inferring mechanisms of
  auditory attentional modulation with deep neural networks,'' \emph{Neural
  Computation}, vol.~34, no.~11, pp. 2273--2293, 2022.

\bibitem{zheng2022general}
Y.~Zheng, H.~Yang, T.~Zhang, J.~Bao, D.~Chen, Y.~Huang, L.~Yuan, D.~Chen,
  M.~Zeng, and F.~Wen, ``General facial representation learning in a
  visual-linguistic manner,'' in \emph{Proceedings of the IEEE/CVF Conference
  on Computer Vision and Pattern Recognition}, 2022, pp. 18\,697--18\,709.

\bibitem{ma2021towards}
P.~Ma, B.~Martinez, S.~Petridis, and M.~Pantic, ``Towards practical lipreading
  with distilled and efficient models,'' in \emph{IEEE International Conference
  on Acoustics, Speech and Signal Processing (ICASSP)}, 2021, pp. 7608--7612.

\end{thebibliography}

\end{document}